\newif \ifdraft
\draftfalse

\documentclass[authorversion,nonacm,acmsmall]{acmart}

\acmJournal{TOSEM}



\usepackage [T1]       {fontenc}
\usepackage [utf8]     {inputenc}
\usepackage [norelnum] {figures}
\usepackage{comment}
\ifdraft
  \usepackage {drafts}
\else
  \usepackage [nodrafts] {drafts}
\fi

\colorlet{draftColor}{orange}

\usepackage[inline]{enumitem}

\usepackage {
  cleveref,
  tabularx,
  tikz
}

\usetikzlibrary [calc]
\Crefname {section} {Section} {Sections}
\crefname {section} {Sect.} {sections}
\Crefname {table} {Table} {Tables}
\crefname {table} {Tbl.} {tables}

\newtheorem {definition} {Definition}

\newcounter {aerq}
\newcommand \aerqid [1]
  {\refstepcounter {aerq}\label {aerq:#1}\theaerq}

\newcommand \cfaerq [1]
  {\textbf{RQ\ref{aerq:#1}}}

\title {Abstraction Engineering}

\author {Nelly Bencomo}
\orcid {0000-0001-6895-1636}
\affiliation {
  \institution {Durham University}
  \department {Computer Science}
  \city {Durham}
  \state {Durham}
  \country {UK}
}
\email {nelly@acm.org}

\author {Jordi Cabot}
\orcid {0000-0003-2418-2489} 
\affiliation {
  \institution {Luxembourg Institute of Science and
                Technology and University of Luxembourg}
  \city {Esch-sur-Alzette}
  \country {Luxembourg}
}
\email {jordi.cabot@list.lu}

\author {Marsha Chechik}
\orcid {0002-6301-3517} 
\affiliation {
  \institution {University of Toronto}
  \department {Department of Computer Science}
  \city {Toronto}
  \state {Ontario}
  \country {Canada}
}
\email {chechik@cs.toronto.edu}

\author {Betty H.~C.~Cheng}
\orcid {0000-0001-9825-5359} 
\affiliation {
  \institution {Michigan State University}
  \department {Department of Computer Science and Engineering}
  \streetaddress {428 S. Shaw Lane, 4115 Engineering Building}
  \city {East Lansing}
  \state {Michigan}
  \postcode {48824}
  \country {USA}
}
\email {chengb@msu.edu}

\author {Benoit Combemale}
\orcid {0000-0002-7104-7848} 
\affiliation {
  \institution {University of Rennes}
  \department {IRISA and Inria}
  \streetaddress {Campus de Beaulieu}
  \city {Rennes}
  \postcode {35042}
  \country {France}
}
\email {benoit.combemale@irisa.fr}

\author {Andrzej Wąsowski}
\orcid {0000-0003-0532-2685} 
\affiliation {
  \institution {IT University of Copenhagen}
  \department {Department of Computer Science}
  \streetaddress {Rued Langgaards Vej 7}
  \city {Copenhagen}
  \postcode {2300}
  \country {Denmark}
}
\email {wasowski@itu.dk}

\author {Steffen Zschaler}
\orcid {0000-0001-9062-6637} 
\affiliation {
  \institution {King's College London}
  \department {Department of Informatics}
  \streetaddress {Bush House, 30 Aldwych}
  \city {London}
  \postcode {WC2B 4BG}
  \country {United Kingdom}
}
\email {szschaler@acm.org}

\begin {document}

\renewcommand \shortauthors
  {Bencomo, Cabot, Chechik, Cheng, Combemale, Wąsowski, Zschaler}

\begin {abstract}

    Modern software-based systems operate under rapidly changing conditions and face ever-increasing uncertainty. 
    In response, systems are increasingly adaptive and reliant on artificial-intelligence methods.
    In addition to the ubiquity of software with respect to users and application areas (e.g., transportation, smart grids, medicine, etc.), these high-impact software systems
    necessarily draw from many disciplines for foundational principles, domain expertise, and workflows. Recent progress with lowering the barrier to entry for coding has led to a broader community of developers, who are not necessarily software engineers. 
    As such, the field of software engineering needs to adapt accordingly and offer new methods to systematically develop high-quality software systems by a broad range of experts and non-experts.
%
    This paper looks at these new challenges and proposes to address them through the lens of \emph{Abstraction}. 
    Abstraction is already used across many disciplines involved in software de\-ve\-lop\-ment---from the time-honored classical deductive reasoning and formal modeling to the inductive reasoning employed by modern data science. The software engineering of the future requires \emph {Abstraction Engineering}---a systematic approach to abstraction \emph{across} the inductive and deductive spaces.  We discuss the foundations of Abstraction Engineering, identify key challenges, highlight the research questions that  help address these challenges, and create a roadmap for future research.
\end {abstract}

\maketitle

\section {Introduction and Motivation}%
\label{sec:introduction}

Abstraction has long been the key to computing\,\cite {ParnasBook2001,Kramer07} and, in particular, to software engineering\,\cite {kramer2006role}. And it remains so. Consider the following example. A part of the central elevator system in a Boston hospital breaks down. Its repair requires to shut a part of the system down for a day.     The elevators are regularly used by patients coming into the hospital for appointments. When an elevator is out of service, these patients are redirected to alternative paths, considering their ability to use stairs. The elevators are also a central transport route for patients moving from a ward to a surgical theatre. These include intensive-care patients, who are transported on mobile beds, which require an elevator. To ensure patient safety, oxygen is available at strategic points along the typical routes. Patients come to the operating theatre from different sources within the hospital: the intensive-care unit is on the same floor as the operating theatres, but the emergency-care department is on the ground floor and needs access to first-floor surgical theatres depending on the needs of incoming patients. Since, the hospital operates close to its capacity, patients may not always have beds available in ``their'' department. Additionally, scheduled procedures also need access to the operating theatres---for example, a planned heart surgery. When planning the elevator repair, Charlene, a clinical lead in this hospital, considers \emph{who} are the different user groups and \emph{how} an elevator shutdown will affect them. Charlene uses a \emph{Digital Twin} system of the hospital to simulate pertinent \emph {what-if} scenarios. The system, used to plan building maintenance, lowering cost and minimizing risks to patients,  is a virtual model capturing the relevant aspects of the domain such as the run-time performance.  To be effective, it necessarily brings together different types of information:%
\begin {enumerate}

	\item \emph {Structural models} that describe where resources are and what paths connect the operating theatres with other wards;

  \item \emph {Process models} describing activities and workflows in different wards, which of them require elevators and at what point;

  \item \emph {Historical data} that is used to predict demand levels for different services on the day of the repair. It comes from electronic health records, with the associated uncertainty;

  \item \emph {Predictive models} used to assess risks to patients  when medical procedures are rescheduled.  These AI models combine historical and real-time data, such as that from wearable sensors.

\end {enumerate}
The above models need to adapt over time according to the structure and utilization of the hospital services. For example, changes are required to react to major context changes due to infection outbreaks, such as COVID-19 or a natural disaster.

A \emph{digital twin} is an IT system mirroring a complex socio-physical system or process\,\cite{Dalibor+22}. The hospital system is but one example.  Others include city-wide systems for analyzing mobility, electricity provision, and pollution patterns\,\cite{schrotter.ea:2020:digital,ivanov.ea:2020}.  Digital twins enable monitoring, simulation, and analysis to support decision making. A \emph{what-if} analysis allows exploring consequences of potential interventions; a \emph{trade-off} analysis helps allocating resources; \emph{design-space exploration} supports planning optimal evolution for the physical system. The social, political, and scientific realities are constantly evolving, and so do the kinds of questions asked and the analyses that need to be performed.  Different stakeholders are interested in analysis at different levels of granularity (e.g., whole-city view vs transportation-system view), and at different-levels of abstraction (coarse grain traffic flows for planning fare systems vs route planning for individual public transport vehicles).
\looseness -1

Engineering such a system efficiently, effectively, safely, and robustly requires a synergistic and carefully choreo\-graph\-ed interaction between a range of disciplines, including structured modeling, operational and behavioral modeling, data management, data science, and machine learning (ML).
Each of these has an individual strong foundation, but their \emph{safe systematic integration} to build complex systems that adapt over time is not yet sufficiently understood.

Digital twins are far from the only example of complex adaptive computing systems, which we begin to build today, and that are likely to grow in relevance and ubiquity in the future.  \emph {Auto\-nomous cars and robots} use machine-learning-based components for object recognition\,\cite{fujiyoshi.ea:2019:deep}, localization and mapping\,\cite {thrun:2005:probabilistic}, motion planning\,\cite {schwarting.ea:2018:planning} and navigation\,\cite {sepulveda.ea:2018}, steering control\,\cite {lefevre.ea:2016:tase}, and high-level decision making\,\cite {DBLP:journals/corr/BojarskiTDFFGJM16,tampuu.ea:2022}.  These components can execute actions that violate safety.  Increasingly many \emph {cyber-physical systems} also include \emph {learning-enabled components}, even in domains such as critical infrastructure and safety management\,\cite {shein:2020:zdnet,belfadil.ea:2024:leveraging,alvi.ea:2023:deep}. Such systems inherently rely on training data, circumventing explicit abstraction. This poses challenges when the operating environment is not reflected well during training~\cite{Enki-TAAS2021}, which is common, and often infeasible to overcome with just data\,\cite {shalevshwartz.ea:2018:formal,Enlil-SEAMS2021}. The key challenge for autonomous learning-enabled cars and robots is trust in their safety\,\cite {shalevshwartz.ea:2018:formal,Enlil-SEAMS2021}. However, learning components often work as black boxes---the opposite of  composable abstractions. As such, the existing safety assurance methods, which depend on abstractions and compositionality, are not directly applicable to the learning enabled systems~\cite{weyns2021towards,Annunaki-TAAS2024}.

Another example are \emph {personal adaptive information services}, which are analytical tools for personal decision-making regarding travel planning, shopping, tax returns, day-to-day economy, remote house control, leisure planning, etc.\,\cite {DBLP:conf/models/MussbacherABBCCCFHHKSSSW14}.
Elements of such systems are already available in various mobile and web apps for monitoring fitness, diet, travel plans, controlling smart homes, along with the corresponding wearables, ubiquitous sensors, and other devices, integrated into overall frameworks such as Google Assistant.\footnote {\url {https://assistant.google.com/}}
The integrated access to generalized and personalized information is envisioned to eventually empower any modern human in personal decision-making in a similar manner as officials, scientists, and engineers use digital twins professionally.
However, for a personal information service to be effective, it needs to learn from the user behaviors; otherwise, interaction and querying will carry an unacceptably high overhead for the user.
For example, the system should learn that the user tends to cook for a larger group of people on weekends, and suggest a recipe and shopping list knowing the content of her fridge and the list of people who RSVPed her calendar invitation, respecting their dietary restrictions.
Solving such problems requires combining on-the-fly explicit symbolic knowledge (e.g. the user's calendar, the scan of objects in the fridge, the model of dietary preferences) with implicit statistical knowledge (e.g. adaptive, learned models of the user's driving and shopping preferences as well as cooking habits).
\looseness -1

\bigskip

\noindent
While software is revolutionizing the modern world, the modern world necessarily requires radical changes in software engineering. All the scenarios described above illustrate how computing-based automation and adaptability are dramatically increasing in volume, complexity, and ubiquity. This blurs not only the line between engineering-time and execution-time\,\cite{10.1145/1882362.1882367}, but also  between software and the real world as both are fusing into a single fabric. Software systems evolve under frequently changing environments, and are expected to handle  ever-increasing uncertainty. These dynamics require accelerated levels of adaptability---indeed, a \emph{temporal} adaptability, i.e., the ability to adapt not only to a fixed space of variable requirements, but also to an emerging chain of changing requirements\,\cite{reqatruntime}, often driven by incoming input data.

A key to addressing these problems is to realize that they all require the creation, manipulation, and maintenance of \emph {abstractions.} These abstractions target different stakeholders who use different tools and techniques. For example, structural models---perhaps provided as CAD models---provide abstractions to be used by civil engineers, building managers, and resource managers, while electronic health records---provided as natural-language text with some structured information components---provide abstractions targeting clinical staff. Nonetheless, building and resource managers need to be able to extract relevant information from electronic health records, just as clinical staff need to understand enough from the structural models to inform their day-to-day work. Equally, the abstractions vary in time: some are defined once and remain static, while others change over time. For example, the core abstractions for describing process models remain static, even if the processes themselves change over time. Many abstractions are no longer intentionally crafted by experts, but are constructed implicitly, i.e., learned or discovered from data, such as via machine-learning techniques, through automated reasoning, or combinations thereof.  Machine learning models yield new abstractions dynamically, while historical data use pre-defined abstractions.

Therefore, it is paramount for the future of software engineering that we broaden and redefine our understanding of abstractions and support their engineering across different disciplines during the whole life cycle.
We call this process  \emph {Abstraction Engineering} (AE). In this article, we discuss the foundations of AE, identify key challenges such as the ones described above, and highlight the AE-related research questions that will help address these challenges.

We proceed as follows. \Cref{sec:abstraction} defines abstraction engineering and gives a taxonomy of modern abstractions, illustrating them in the context of the motivating scenarios described above. \Cref{sec:challenges} discusses the key challenges faced by engineers developing contemporary systems and uses these to highlight open research questions in abstraction engineering. We survey some partial answers to these questions in \cref{sec:solutions}, and discuss the further research looking forward in \cref{sec:conclusion}.

\begin {figure*} [t]

\def \features {
  struct/4.4,
    det/7.17,
    stoch/7.05,
  ui/5.1,
  ai/4.4,
  onto/5.1,
  dots/3.5,
    expert/7.9,
    reqEng/7.9,
    sysDesigner/6.9,
    dataScientist/7.7,
    sysDeveloper/7.9,
    qAEng/7.9,
    user/6.7,
  machine/4.5,
    elicitation/7.3,
    design/7.4,
    implementation/7.4,
    qa/7.4,
    analysis/7.4,
    dynUpdated/7.15,
    dynCreated/7.1,
  topDown/4.9,
  bottomUp/5.2
}

\begin {tikzpicture} [
  every node/.style = {black, anchor = north west},
  every path/.style = {very thin, gray}
]

  \newcommand \colWidth {0.423 \textwidth}

  \newcommand \titleY {3.4}

  \def \rowOffset      { 0.200} 
  \def \rowSkip        {-0.392} 
  \def \rowOffsetLabel { 3.00} 
  \def \rowSkipLabel   {-0.93} 

  \coordinate (labelhead) at (7.87, \titleY);
  \coordinate (bendhead)  at (7.3, 0);

  \newcommand \exHeader [1] {\textsf {\textbf {\strut \small #1}}}

  \newcommand \ftLabel  [3][0mm]
    {\node (l#2) [
        text width = \colWidth,
        align = justify,
        font = \small%
               \setlength \baselineskip {.8 \baselineskip},
        anchor = west,
        inner ysep = 0mm,
        yshift = #1,
        ] at (label#2) {\textls[-5]{\footnotesize\strut #3}};

      \draw [very thin, gray] (l#2.west) -- (bend#2) -- (#2);
      \draw [very thin, gray]
        (l#2.south west) -- (l#2.north west) -- ++(right:1mm);
    }

  \node [
    inner sep = 0mm,
    outer sep = 0mm,
    overlay
  ] at (0, 0) {
    \includegraphics [
      width = .55 \textwidth,
      draft = false
    ] {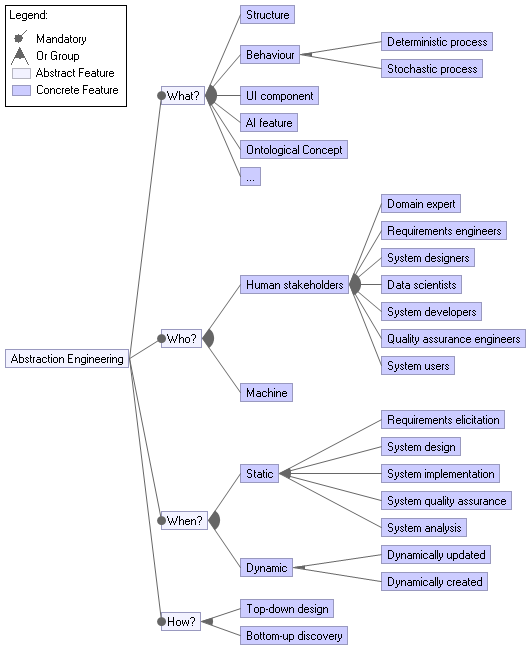}
  };

  \foreach \Feature/\Column [count = \ind] in \features {

    \pgfmathsetmacro \yCorner
      {\ind*\rowSkip + \rowOffset}
    \pgfmathsetmacro \yLabel
      {\ind*\rowSkipLabel + \rowOffsetLabel}

    \coordinate (\Feature)  at (\Column, \yCorner);
    \coordinate (y\Feature) at (\Column, \yLabel);
    \coordinate (bend\Feature)  at (\Feature  -| bendhead);
    \coordinate (label\Feature) at (y\Feature -| labelhead);

  }

  \node [text width = \colWidth, inner sep = 0mm] at (0.0, \titleY)
    {\exHeader {\rlap{A Taxonomy of Abstraction Engineering}}};

  \node [text width = \colWidth, inner sep = 0mm] at (labelhead)
    {\exHeader {Selected Examples in Contemporary and Future Systems}};

  \ftLabel {struct}
    {Define structure and routes within a hospital, schema of data sources for personal information systems};

    \ftLabel [1mm] {det}
    {Models \emph {of} the digital-twin system itself support its engineering};

  \ftLabel {stoch}
    {Models of service process in a hospital, a model of pedestrian localization module, reliability characteristics for sensors are probabilistic};

  \ftLabel [9mm] {ai}
    {ML models can adapt a driving style or efficiency of control for a specific vehicle and engine};

  \ftLabel [10mm] {onto}
    {World models (knowledge base) of an autonomous robot are often expressed using ontologies};

  \ftLabel [15mm] {expert}
    {Domain experts, designers, system and safety engineers work on digital twin and autonomous system projects as in traditional software engineering. The advent of ML has raised the barrier to contribute for these experts. Data scientist create abstractions from historical and real-time data
    \looseness -1};

    \draw (dataScientist) -- ++(up:13mm) -- ++(170:1mm);
    \draw (dataScientist) -- ++(down:9.5mm) -- ++(190:1mm);

  \ftLabel [50mm] {user}
    {Users of cars and personal information systems define preferences and priorities for the service.  Digital twin users have to implement complex rule changes in the system. Smart interfaces (like language models) are needed to support users creating these abstractions};

  \ftLabel [45mm] {machine}
    {Learning algorithms fit predictive models to improve short term planning (robots and autonomous cars) and to improve predictions (digital twins)};

  \ftLabel [43mm] {elicitation}
    {Static abstractions, elicited in design and implementation will remain used in safety-critical systems for assurance purposes (cars, robots)};

  \ftLabel [77mm] {dynUpdated}
    {Static abstractions used for certification will be monitored at runtime to control operational conditions and to accumulate assurance evidence};

  \ftLabel [75mm] {dynCreated}
    {New abstractions needed for long living digital twins, and for highly-personalized information systems, created as new use cases arise};

  \ftLabel [80mm] {bottomUp}
   {A world model of an autonomous robot is build using input from the perception stack. New kinds of facts can be extracted from electronic patient records using language models as hospital procedures change
   \looseness -1};

\end {tikzpicture}

\caption {A deceivingly simple feature model of types of abstractions. A different configuration of the feature model may apply to each individual abstraction.  The right column of the figure underlies how the changing characteristics of software systems, changes the nature of modern abstractions and their engineering.}%
\label {fig:questions_fm}

\end {figure*}
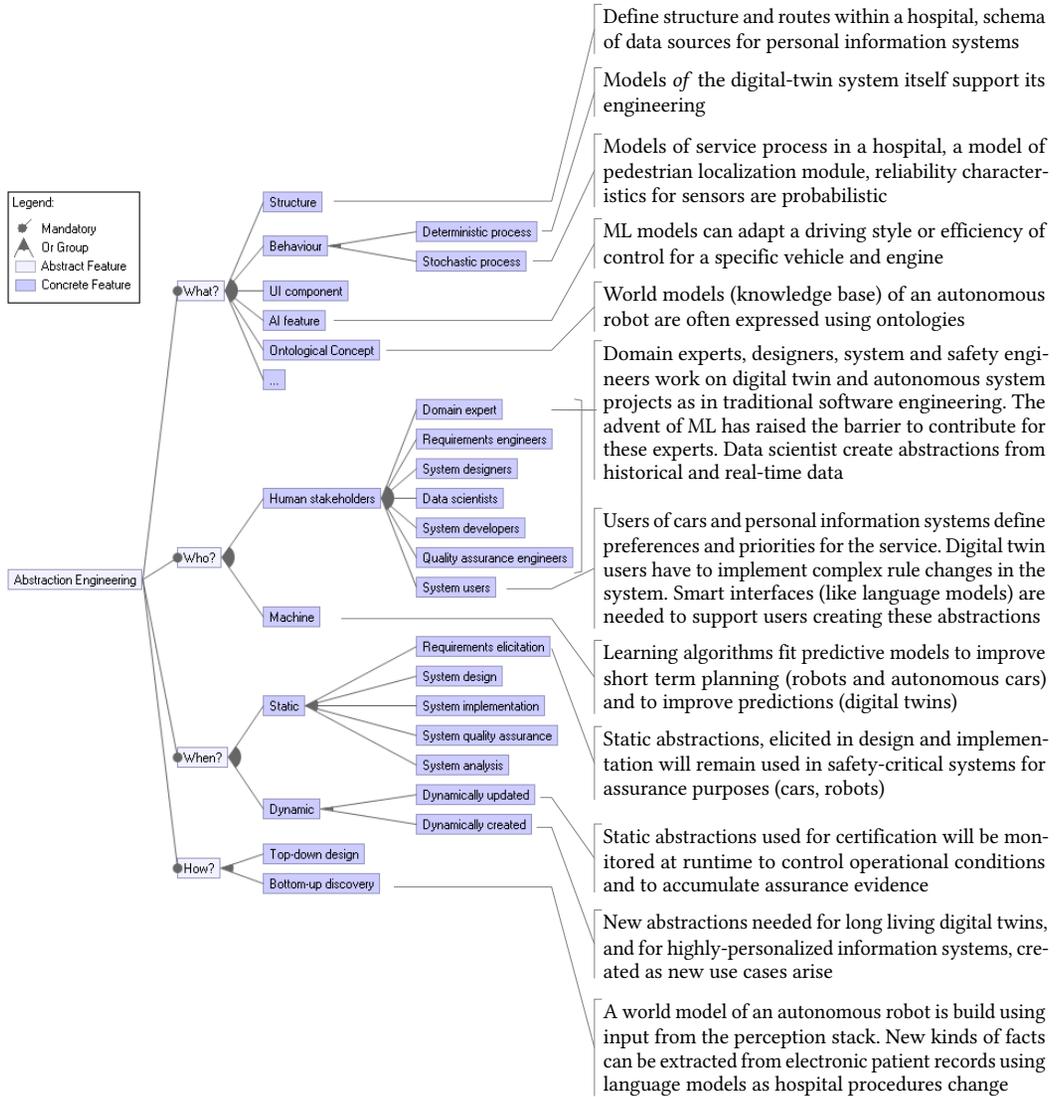

\section {Abstraction Engineering}%
\label {sec:abstraction}

In this section, we define the key terms \emph {abstraction} and \emph {abstraction engineering}. Next, we characterize different types of abstractions along four dimensions.

\subsection {Foundations}

Abstractions have been identified as a crucial skill for software engineering professionals since the early ages of computing---a means by which engineers deal with complexity, including both removing detail and identifying generalizations\,\cite {kramer2006role}. Orthogonally to this process-oriented perspective, some authors consider abstractions as things, objects of interests, and manipulatable entities\,\cite {Kiczales1991,Visser15}. We follow the latter school of thought:
\begin {definition}

  An \emph {abstraction} is a representation of a concept of concern in a particular context. Each abstraction, as a minimum, has a name and a purpose or intention.

\end {definition}

\noindent
We strongly believe that the process of engineering such abstractions will be key to the future of  modern software engineering. We define abstraction engineering as follows:

\begin {definition}

 \emph {Abstraction Engineering} (AE) is the discipline of constructing and manipulating \emph {abstractions} for a given purpose.

\end {definition}

\noindent
Abstractions are used in many areas of computing under different names: for example, features in machine learning, classes in object-oriented source code, types in functional programming languages, concepts in ontologies, variation points in product line engineering.  Engineers construct abstractions for many purposes---to understand, capture, communicate, and manage domain knowledge; to support reasoning; to structure systems and development processes; and to describe the essence of the problem at hand.

While all software engineers use abstractions, abstraction engineering makes abstractions explicit---giving them a first-class status. Abstraction Engineering goes beyond a pure mathematical notion of abstraction (e.g., as in abstract data types) and is concerned  with both construction of abstractions and with manipulating them.
\begin {description}

  \item [Construction\textnormal,] whether performed by humans or algorithms, requires methods and tools for identifying, specifying, validating, and evolving abstractions. This includes support for managing collections of inter-related abstractions and patterns for their use. To predictably engineer  high-quality abstractions, system designs, and, ultimately, the systems, the construction process must be systematic and repeatable.

  \item [Manipulating] abstractions includes using them prescriptively, descriptively, or predictively\,\cite {Combemale+21DS} to construct (develop, generate, execute, or simulate) systems and to support the system design and deployment process overall.
  Also included are prediction, classification, evaluation, testing, verification \& validation, run-time monitoring, decision making, explanation, generation of documentation, and etc. To meet the contemporary requirements for temporal adaptability, manipulation must be easy, even agile, while maintaining the high quality of the abstractions. This can challenge both human engineers, who face increasing system complexity and increasing pace of changes, as well as machine learning methods, which suffer from model deterioration in incremental learning when distributions drift.

\end {description}

\subsection {Anatomy of an Abstraction}%
\label{sec:anatomy}

To better understand Abstraction Engineering, we look at it through the lens of four characterizing questions: \emph {`What?', `Who?', `When?'}, and \emph {`How?'}.  \Cref {fig:questions_fm} gives a high-level overview of these perspectives using a feature model---itself an example of an Abstraction Engineering method from the software product-line community\,\cite {Kang+90}.  We discuss each question in detail, showing how abstractions from the examples of the previous section are characterized.

\paragraph {What is being abstracted? What abstractions capture the functional and non-functional requirements of the system?}
In autonomous cars and robots, abstractions capture, among others, the intended behavior of the learning-enabled components, in a sound manner, so that they can enable analysis and building safety assurance cases for the autonomous machines. For instance, a probabilistic model may abstract a vision module localizing pedestrians under realistic angles, visibility and limited motion velocity. A simple static failure model can be a confusion matrix for the detector success and an error distribution for localization. A more complex model can be a Partially Observable Markov Decision Process (POMDP), able to capture state-based behavioral changes. Furthermore, abstractions can be used to capture assurance objectives, strategies for achieving these objectives, and evidence that these objectives have been met~\cite {Wei+19,MODELS2022-MODALAS-MDE-Assurance-LearningBasedSystems,AC-ROS-MODELS2020}. In digital twins, the abstractions largely depend on the subject area and are of two kinds: models \emph {of} the digital twin, and models \emph {in} the digital twin\,\cite {gray.rumpe:2022}. The former concern the dynamics of the system, i.e., the processes measuring and reacting to measurements in the reality and in the simulation.  The latter concern the actual sensor measurements and actions of agents monitored by the digital twin. For personal information systems, one needs abstractions allowing integration of new data sources, both explicit (e.g., databases) and implicit (e.g., outputs of machine learning models).  Moreover, in the case of machine learning, abstractions need to allow the machine learning models to integrate explicit data (e.g., a calendar) into the training and re-training process of neural networks, to allow enriching the system with new knowledge as it becomes available.

\paragraph {Who creates the abstractions?}

In robotics and autonomous driving, the abstractions are created by requirements engineers, domain experts, and quality assurance engineers.  The same happens in the domain of digital twins, when experts need to describe the processes, and what data is collected. However,  parts of the process dynamics and relevance of the measurements may be learned\,\cite {bennaceur19}, so machine learning algorithms can participate in creating these abstractions. For instance, a learned predictive model can be included. When sophisticated abstract or cross-cutting queries need to be asked, or unexpected analyses are required, the users of the system may need to be able to define the suitable views within the system. User-friendly technology for devising such abstractions is desirable~\cite {bennaceur19}.  For example, personal information systems may allow end-users to integrate new data sources into the service.  This requires an abstraction definition mechanism that end-users could use with low overhead, for instance, based on sketching or natural language.
\looseness -1

\paragraph {When in the software life-cycle is the abstraction defined and used? Is it static throughout the life of the system, or does it change over time?}

For reasons of safety, in robotic and driving systems, the re\-quire\-ments-based abstractions typically do not change at run time. The requirements are usually specified statically, and so are the corresponding abstractions. The safety-assurance--related processes, models, and specifications are also executed and analyzed during the quality assurance process, at design and implementation time.  However, the abstractions partaking in the safety implementation may need to be monitored at run time.  In the autonomous systems industry, assurance cases are also updated based on the user experience,  for instance, past mileage of a particular vehicle design or component.  New methods are emerging to allow updating assurance evidence based on the new  incoming data, which requires monitoring abstractions at run time~\cite {BencomoGS19,calinescu17,MODELS2022-MODALAS-MDE-Assurance-LearningBasedSystems,AC-ROS-MODELS2020}.

A digital twin  can be a long-lived system, existing alongside the real system which it is shadowing. One typically starts with an initial abstraction that must evolve over time as the real system also evolves.  The new analysis and simulation questions arising during the long-term operation also pose requirements to be able to synthesize new views (abstractions) on the system.  The need for long-term adaptation is present in many digital twin systems, but perhaps is most pronounced for personal assistance services which need to adapt to \emph {ad hoc} wishes of end users.

\paragraph {How is the abstraction designed?  Is it built in a top-down fashion or discovered bottom-up from data?}

For cars and robots, the key abstractions capture observable behavior and require domain experts to apply a \emph {top-down} approach to create the appropriate models. Some abstractions may be created through hybrid and adaptive methods, where a human initially defines an abstraction that is then refined \emph {bottom-up} from data. Example of a common case is the \emph {world model} of an autonomous system.  Typically, experts design the language of these models (known as a \emph {meta-model}), while the system uses sensor measurements and learning-enabled components to populate this model bottom-up with object instances (pedestrians, traffic signs, road layout, other vehicles)\,\cite {tenorth.beetz:2009:knowrob}. Abstractions in a digital twin are both created and discovered. Indeed, some abstractions are created with the help of domain experts that decide what elements of the real-world objects are important for the twin's purpose and therefore need to be part of the design. Others are discovered when connecting to the sources of data whose schema and contents are dynamic, such as electronic patient records. For personal assistance systems, we would like to have increasingly many discovered abstractions, to account for the high diversity among users, and the high pace of change in their life habits.

\newcommand
  \tblHeader [1]
  {\textbf {\footnotesize #1}}

\newcommand \zipCell [1] {\textls[-11]{#1}}

\begin{table*}[!p]

\vspace{-2mm}

\begin {tabularx} {\textwidth} {
  @{}
  >{\footnotesize}r
  @{\hspace {0.4mm}}
  >{\footnotesize}X
  @{\hspace {1.7mm}}
  >{\looseness -1\raggedright\footnotesize}p{24mm}
  @{\hspace {1.3mm}}
  >{\looseness -1\raggedright\footnotesize}p{25mm}
  @{\hspace {1.3mm}}
  >{\looseness -1\raggedright\footnotesize\arraybackslash}p{24mm}
  @{}
}

  & \tblHeader {~\break Research Question}
  & \tblHeader {~\break Digital Twins}
  & \tblHeader {Autonomous Cars and Robots}
  & \tblHeader {Personal Information System}
  \\[-0.8mm]
  \midrule

  \aerqid {dynamic_abstractions}
  & \zipCell{How to robustly manage abstractions that are dynamically discovered while a system runs, continually producing and collecting data?} 
  & Efficiency models for hospital processes based on data
  & Specification for perception stack
  &
  \\

  \aerqid{different_sources}
  & \zipCell{What are the differences between analyses for \emph{designed} vs \emph{discovered} abstractions? Can they be safely and meaningfully combined to reason about composed systems? Is it possible to make machine-learned abstractions more similar to the human-created ones?}
  & \zipCell{Models in digital twins combine designed abstractions with data from sensors}
  & \zipCell{Safety assurance combines abstractions of code and ML-components}
  &
  \\

  \aerqid{context_choice}
  & \zipCell{How to represent an abstraction context and select abstractions for a given use context? How to discover new abstractions for new contexts automatically?}
  & \zipCell{Different use cases in the hospital need different abstractions}
  & \zipCell{Expand autonomous driving to new cities and jurisdictions}
  & \zipCell{User contexts and use cases are constantly evolving}
  \\

  \aerqid{stakeholders}
  & \zipCell{How to align and synchronously maintain abstractions developed for and by different stakeholders, from different perspectives? What are the agreements and conflicts between different simultaneous abstractions,  how can they be reconciled systematically?}
  & \zipCell{Different stake\-hol\-ders in the hospital need different abstractions}
  & \zipCell{Road management and car users need different abstractions}
  & \zipCell{Users and service providers use different abstractions to describe the same service}
  \\

  \aerqid{specificity}
  & \zipCell{How do we manage abstractions of varying specificity from the very precise to the very uncertain? How do we compose information of different uncertainty to assess the certainty of system analysis?}
  & \zipCell{Fixed\,wards,\,routes; uncertain number and kind of medical procedures}
  & \zipCell{Sensor and ML-model perf.\ specs uncertain; driving code rules precise}
  & \zipCell{Precise tax rules; uncertain return specification for investment products}
  \\

  \aerqid{uncertainty}
  & \zipCell{How to assess and manage the uncertainty inherent in abstractions? What is the trade-off between uncertainty and the capability of reasoning at scale, facing compounding uncertainty and information loss?}
  & \zipCell{Data entry about patients relies on overworked staff}
  & \zipCell{Vehicle state is only estimated, as sensors are few and imprecise}
  & \zipCell{Data source may be offline---reason with available or outdated data}
  \\

  \aerqid{aleatoric_uncertainty}
  & \zipCell{How do we control and reason with aleatoric uncertainty that ``pollutes'' abstractions learned from collected data? How can we specify correct behavior under aleatoric uncertainty?}
  & \zipCell{Biological processes are complex, making predictions of risk difficult}
  & \zipCell{What does it mean to drive safely if actions of pedestrians are unpredictable?}
  &
  \\

  \aerqid{epistemic_uncertainty}
  & \zipCell{How to assess and reduce epistemic uncertainty of abstractions, especially for models with temporal dependencies, and embedded in complex control loops?}
  &
  & \zipCell{System-level epistemic uncertainty is not well-defined for a vehicle}
  & \zipCell{Complexity of open world makes scoping models difficult}
  \\[-1pt]

  \aerqid{reason_at_scale}
  & \zipCell{How do we  effectively compose many sources of uncertainty with open world variability of ever expanding technologies, models and extension mechanisms}
  & \zipCell{New medical devices interact with existing processes in unforeseen ways}
  & \zipCell{New communication protocols, apps, devices supported add to uncertainty}
  &
  \\

  \aerqid{composing}
  & \zipCell{What are composition properties of AI and non-AI components? How can we decompose them and recompose them? What properties are preserved then?}
  & \zipCell{How to combine predictions from different models of hospital load?}
  &
  & \zipCell{Combine\,weather forecast,\,news, ticket prices to choose the  way to travel}
  \\[-3pt]

  \aerqid{abstraction_relationships}
  & \zipCell{What relationships exist between abstractions (e.g., consistency and derivation relationships)? How can we derive them and reason about them?}
  & \zipCell{Combine a digital twin of patients inflow with a twin of COVID proliferation in the area}
  & \zipCell{Perception\,perf.\,in darkness and in rain do not transfer to simultaneous rain and darkness}
  &
  \\[-3pt]

  \aerqid{pragmatics}
  & \zipCell{What are the pragmatics of how abstractions are used? How do the uses of specific abstractions shift over time and what are the implications for reasoning?}
  & \zipCell{Same models are used to answer new questions}
  & \zipCell{Build adaptive\,cruise control using obstacle distance from emergency braking}
  &
  \\[-3pt]

  \aerqid{hierarchies}
  & \zipCell{What instantiation mechanisms support reuse of abstractions across levels of detail to systematically create new abstractions? How to derive abstractions reusable across highly sophisticated particular applications? Higher abstraction levels needed?}
  & \zipCell{Digital twins are alike at high abstraction levels, but their implementations are bespoke}
  & \zipCell{Capture recurring patterns of argumentation as abstractions in their own right?}
  & \zipCell{Abstract\,from user needs in the implementation yet support extreme runtime customization}
  \\

\end{tabularx}

\caption {Abstraction Engineering research questions. The rightmost three columns link the questions to our example domains, illustrating the type of cases for which the questions could be studied.}%
\label {table:aerq}

\end{table*}

\section {Key Challenges}%
\label{sec:challenges}

Having analyzed the three domain examples and the different kinds of abstraction engineering processes, we look into the key challenges ahead of us---namely, \emph{complexity} and \emph{uncertainty} associated with the Problem Space (i.e. the problem context where the solution can be applied), and \emph{compositionality} and \emph{reuse} associated with the Solution Space (i.e. the specific technologies, frameworks, and tools used for the solution being applied to the problem). We discuss each of these in turn and illustrate them briefly in the context of the problem domains.

\subsection {Complexity (Problem Space)}

The complexity in modern systems has two main sources: complex, often socio-technical environments, on which these systems are deployed, and interactions and non-linearities within the system itself.  For example, the behavior of machine-learning-based systems, especially\ deep-learning-based, is not directly amenable to analysis as it emerges from the complex interactions between the training data and complex neural networks. Thus, emergent behaviors and properties are often not predictable at the time a system is designed. To identify and manage them, we need to be able to manage abstractions that are dynamically discovered while a system runs\,\cite{BencomoGS19}, producing and collecting data continually (cf.\ \cfaerq{dynamic_abstractions} in \cref{table:aerq}).
\looseness -1

Small changes to design or training can have significant effects---this is characteristic of complex systems. Consequently, these systems defy traditional formal analysis methods, which are often based on hierarchical breakdown. Even the notion of ``correctness'' as an absolute, rather than contingent, term may not be applicable. Yet, AI-based systems, such as autonomous vehicles, use AI to make decisions in the real world (e.g., to navigate or to avoid accidents).  Therefore, AI provides a safety-relevant function in these systems, and we should apply safety assurance methods to it. If an AI-based system malfunctions (e.g., if an autonomous system crashes), how can we understand the causes and prevent future malfunctions? Thus, we need both to understand the different approaches for analysing abstractions whether they are \emph{designed} or \emph{discovered} by a \emph{human} or a \emph{machine}, and to understand how different approaches can be meaningfully combined together (cf.\ \cfaerq{different_sources} in \cref{table:aerq}).

Similarly, personal adaptive information services need to draw on the essentially unbounded knowledge in the user's life and make decisions about what is most relevant to be presented at a given point in time and how to do so. This requires the ability to cope with the complexity of social contexts, where small changes can have profound implications, and it requires the service to provide only relevant information, in the right form, and that can be understood and used by its user in the current situation.
To provide such services, we need to understand abstractions as contextualized objects, represent the context of an abstraction and select the right abstraction based on the context (cf.\ \cfaerq{context_choice} in \cref{table:aerq}).

Digital twins present views of complex real-world systems, at different levels of granularity, over different time horizons, and for different stakeholders. They should support analysis of the outcomes of interventions, even for complex system dynamics.  The information needs to be presented in an understandable manner, and achieve trust of  non-technical users, who many not have software expertise.  Furthermore, not all users have intricate understanding of the system dynamics. Different stakeholders need different abstractions, so the challenge is to align and combine abstractions developed and used by different stakeholders, representing different perspectives on the same system or process (cf.\ \cfaerq{stakeholders} in \cref{table:aerq}).
\looseness -1

\subsection {Uncertainty (Problem Space)}

Modern systems manage levels of uncertainty at design time and run time, including specification, environment, and model uncertainty.

\paragraph {Specification uncertainty.}

For AI-based systems, even the specification of what constitutes a ``valid'' result may be vague.  A perception component in a self-driving car is often charged with ``detecting a pedestrian''---but what \emph {is} a pedestrian, precisely?   And does the pedestrian need to be detected when there is very poor visibility or the windshield of the car is fully covered by frost\,\cite {hu22}?  While  AI-based systems are often employed specifically to manage such uncertainty, if even the specification of what constitutes a hazard is uncertain, how can we successfully argue for system safety? We need to understand how to manage abstractions of varying specificity from the very precise to very uncertain, and how to combine the information to assess the certainty of the analysis of a system (cf.\ \cfaerq{specificity} in \cref{table:aerq}).

\paragraph{Environment uncertainty.}

Systems are put into environments that expose high levels of uncertainty. At design time, it is impossible to accurately capture all possible scenarios that a system may encounter. At run time, a system needs to cope with \emph{measurement uncertainty} from its sensors. For example, a digital twin of healthcare operations  relies,  at least in part, on manual data entry by healthcare staff, who may be overloaded and may prioritise working with patients over data capture, potentially affecting timeliness and level of detail of information captured. We need to understand how to assess and manage the measurement uncertainty inherent in all abstractions, and how to balance this against the capability of reasoning at scale due to appropriate levels of information loss in abstraction (cf.\ \cfaerq{uncertainty} in \cref{table:aerq}).

The environment is often a source of the \emph{aleatoric uncertainty}, i.e., phenomena that we have to deal with but are inherently uncertain, for instance, actions of pedestrians and other drivers, or contradictory markings of lanes on the road due to a recent lane reorganization.  We need to recognize that aleatoric uncertainty is inherent to some automatically derived abstractions (cf.\ \cfaerq{aleatoric_uncertainty} in \cref{table:aerq}).
\looseness -1

\paragraph{Model uncertainty (epistemic).}

ML models are trained on finite sets of data points and then extrapolate outputs for new inputs.  Thus, ML models are inherently approximating: embedding uncertainty, known as epistemic, in every output. During operation, we need to know whether the system is still being used within the validity envelope of the training data, or whether the operational distribution has drifted. This is particularly difficult for models with a dynamic temporal component, with memory, recurrence, or embedded in a control loop, making the output dependent on the past. New ways to quantify and reduce the uncertainty of models  are needed, for both created and discovered abstractions (cf.\,\cfaerq{epistemic_uncertainty} in \cref{table:aerq}).
\looseness -1

High configurability of modern systems adds uncertainty, as it is impossible to predict all possible configurations at design time and, therefore, all possible feature interactions.  For example, drivers or passengers connect mobile phones to a car using a range of communication channels, creating a potential feature interaction not only with the infotainment system but also with safety-relevant features, such as navigation. The problem of uncertainty is exacerbated when we compose systems from multiple AI components, where we then need to address the problem of the corresponding composition of uncertainty\,\cite{DBLP:journals/sosym/CamaraTVBCCGS22}. How do we  effectively compose many sources of uncertainty with open world variability of ever expanding technologies, models and extension mechanisms (cf.\,\cfaerq{reason_at_scale} in \cref{table:aerq})?
\looseness -1

\subsection{Compositionality (Solution Space)}

\emph{Compositionality} and \emph{modularity} are fundamental principles of systematic engineering. They allow the development of large systems to be broken down into independent sub-tasks.  Yet modern systems are challenging to develop in such a modular fashion. For example, we may want to use multiple AI classifiers and base decisions on a majority vote between them to mitigate shortcomings of individual classifiers. Alternatively, we may require different AI components to complete different tasks, for example, a recognition component to identify objects in the street and a planning component to derive adaptations to the current navigation plan. However, composing different types of AI models this way is not straightforward and requires significant expertise in each method to ensure that the results can be trusted.  (cf.\ \cfaerq{composing} in \cref{table:aerq})

Not only is it difficult to compose AI models,  we also cannot compose any evidence that captures  assurance properties about a given AI-based component. For example, suppose that an obstacle detection subsystem has been assessed to work robustly in the rain. We cannot easily combine that assessment with the one that the component works well in low light to claim that it works well in dark and rainy conditions\,\cite{hu23}.  To make the assessment, we need to reason about the relationships between abstractions, in particular, how one is coordinated with, or derived from another (e.g., how abstractions about the presence of objects are derived from abstractions about the composition of images perceived by the sensors (cf.\, \cfaerq{abstraction_relationships} in \cref{table:aerq}). Similar concerns apply to the composition of digital twins: it is difficult to establish the validity frame for a digital twin resulting from the composition of two or more digital twins even where their respective validity frames are known, because validity is contingent on context and environment, which are affected by the composition.
\looseness -1

The task of composing AI-based and traditional sources of information, such as in personalized adaptive information systems or in digital twins, is even more challenging. Combining explicit knowledge with implicit knowledge learned by AI systems, possibly deciding which should take priority or adapt the other, is currently an unsolved challenge.  There is also tension between different priorities: for instance, what is more important, your diet or the general climate data? Profoundly, in logics, knowledge is naturally composable; however, knowledge in information-theoretical sense, like in machine learning, is difficult to compose. We need to develop strategies for combining \emph{discovered} and \emph{engineered} abstractions~\cite{Bencomo2013}. We also need to understand relationships of agreement and conflict between abstractions used by different stakeholders and how to reconcile them systematically (cf. \cfaerq{different_sources} and \cfaerq{stakeholders} in Table~\ref{table:aerq}).

\subsection {Reuse (Solution Space)}
The ability to reuse existing components in new contexts is essential to the effective development of software systems. It is an important engineering principle and a sign of a maturity of a discipline. However, reuse in modern systems  is challenging. For example, once an AI-based system has been developed and  its quality assured, AI components cannot be directly reused in new contexts. This is because it is unclear which quality assurance outcomes can be transferred to the new context and which assurance activities have to be repeated (see for instance \cfaerq {composing}).

Digital twins embed different models to support  services provided to the end-users. A model can play different roles, possibly over time, to support different services\,\cite {DTConcepts}. In particular, the representation of the system requires descriptive models, to be used in combination with predictive models for conducting predictive simulation, and possibly with prescriptive models to act back on the system. While some models can be borrowed from design time, others need to be defined and built at run time. We need to understand the pragmatics of how abstractions are used and how their uses may shift over time and to new contexts (cf.\,\cfaerq {pragmatics} in \cref{table:aerq}).

Moreover, while there has been an explosion of digital twins (for mobility, for smart grid management, for industrial plants, for water monitoring, etc), each of them is developed from scratch, even though all digital twins share a set of common building blocks.  That is, the development of digital twins is artisanal, or ad hoc, rather than industrial~\cite{Niederer2021}. There is a need for hierarchies of abstractions that can be instantiated in diverse ways (e.g., as strict templates or as more flexible patterns) to systematically create new abstractions  for particular applications (cf.\,\cfaerq {hierarchies} in \cref{table:aerq}).

Personalized adaptive information systems would also benefit from reuse ``at scale'': every personalized instance is aiming to reuse existing components in a new context. We do not currently know how to enable such reuse in an automated fashion. We need to work out how to build systems that abstract from an individual user in the implementation, but are highly adaptive and idiosyncratic at run time.  Again, this requires an understanding of different ways in which abstractions can be instantiated. Here, for example, created abstractions (describing people ``in general'') would be used as `guard rails' to guide the discovery of abstractions specific to an individual user based on data collected about them (cf.\,\cfaerq{stakeholders} in \cref{table:aerq}).
\looseness -1
\section {A Medley of Partial Solutions}%
\label {sec:solutions}

Many of the research questions identified in \cref {sec:challenges} have (partial or full) solutions developed in particular fields of Computer Science, making them difficult to apply in other contexts.  For example, uncertainty challenges are addressed in \emph{formal logics} such as subjective logic\,\cite {Josang01}, 
testing~\cite{gerhold2018model,wang.ea:2023:fse,DBLP:journals/pacmpl/VarshosazGJW23,AdaptiveTesting-SEAMS2014}, verification~\cite{prism,albarghouthi12,dippolito08}, and requirements engineering~\cite{ISO21448:2022,DBLP:journals/re/SalayCHS13} for systems displaying probabilistic behaviors or operating in uncertain conditions.

\emph{Language-oriented programming}~\cite {Ward1994} is an approach to software development that explicitly creates new language constructs (\emph{abstractions}) as part of the programming effort.
Building on this, \emph{model-driven engineering}~\cite{Brambilla+17} and \emph{software language engineering}~\cite {10.5555/1496375} focus on engineering tool-supported abstractions in so-called domain-specific (modeling) languages~\cite {mdebook,dsldesign}, but, on their own, do not answer the question of what should be abstracted or how to manage abstractions generated from data or by AI. In this context, language reuse~\cite {butting2020compositional} has been studied as a way of reusing abstractions from one domain to another. Similarly, language composition operators~\cite {10.1145/2427048.2427055} are used to align and complement abstractions from different domains.

Abstraction has been a key concept in the verification community, from defining automated reasoning over useful ``fixed'' abstractions via \emph{abstract interpretation}~\cite {CousotCousot77} to dynamically discovering abstractions through \emph{counter-example-based abstraction refinement (CEGAR)} approaches to software verification~\cite {Clarke+00}. The goal was less about managing uncertainty or enabling reuse but about defining abstractions that are small enough to enable decision procedures and yet precise enough so that analysis results can be meaningful.

AI can help with the abstraction  process. AI technology is already contributing to many software engineering tasks~\cite {10.1145/3505243,10.1145/3485275} including requirements engineering and software specification and design where abstractions play an important role. AI could help transform natural language descriptions (e.g., coming from interviews with clients, organizational practices, operation manuals, etc.) into models that capture the key elements of these descriptions and abstract the rest.  We are starting to see the application of LLMs to modeling tasks~\cite {camara2023assessment,fill2023conceptual} even if, so far, the work is more focused on exploring opportunities and limitations. Machine learning techniques are also used to create smaller and more explanatory abstractions of large and complex machine learning models\,\cite {viper}.

Foundational ontologies (e.g., \cite {Guizzardi05}) aim to identify general categories of concepts (\emph{abstractions}) that can be applied across conceptual models for different domains. These foundational abstractions can help align different models that use similarly named concepts, by enabling an analysis of whether these concepts do indeed refer to the same real-world object. Though, while foundational ontologies have been explored in the context of conceptual modelling, it is less obvious how they apply to data science or behavioral modelling.

\section{Roadmap}
\label{s:roadmap}
    
  Our ability, as a community, to construct quality software in the age of AI critically depends on our success in establishing abstraction engineering as a principled lens on software engineering. We discuss the approaches to this challenge from four different perspectives: 
  \begin{enumerate*}[label={(\alph*)}, ref={(\alph*)}), itemjoin={;\ }, itemjoin*={; and\ }, afterlabel={~}]
    \item technical research challenges
    \item education
    \item artifacts to be produced
    \item ways of working
  \end{enumerate*}%
  .
    
  \subsection{Technical research challenges}

    Abstraction Engineering will build on techniques from several software engineering areas 
    (model-driven engineering, modeling \& simulation, formal methods, AI, etc.), but will define new research challenges for each individual discipline and for their combination as a whole.
    These research challenges will fall into two categories: 
    \begin{enumerate*}[label={(\roman*)}, ref={(\roman*)}), itemjoin={;\ }, itemjoin*={; and\ }, afterlabel={~}]
      \item \emph{foundational Abstraction Engineering research} that aims to understand the fundamental characteristics of abstractions and their construction and manipulation (the dimensions implied by the characteristic questions in~\cref{sec:anatomy} are an example here)
      \item \emph{applied Abstraction Engineering research} that aims to understand how Abstraction Engineering approaches developed in different sub-fields of computer science can be combined or transferred to other sub-fields.
            Many of the AE research questions in \cref {table:aerq} fall into this category.
    \end{enumerate*}

  \subsection{Education}
  \label{s:roadmap:education}
  
    
    Changing the perspective of engineers and researchers and establishing Abstraction Engineering as a new discipline will require changes to education.
    We believe that these changes need to focus on the following three areas:
    \begin{description}
      \item[Who do we teach?] 
            Of course, need to teach abstraction engineering to students, especially at the university level,  to enable them to take this broader perspective and see the connections between the use of abstraction in different disciplines.
            In particular, students in the newly emerging AI and data science degrees need to be explicitly taught about engineered abstractions and their relationship to the learned abstractions that can be extracted inductively from data.
            In addition, to really achieve a perspective change across software development, we need to provide training (tutorials, hands-on materials, etc) to practitioners who are already working in the field. 
            Again, we need to do this across disciplinary boundaries, focusing not only on software engineers, but also data scientists, AI engineers, software managers, etc.
      \item[What do we teach?] 
            Abstraction engineering is about a change of perspective, viewing techniques and approaches in different disciplines through the principled lens of \emph{abstraction}.
            Most importantly, therefore, we need to teach how to recognise different approaches to abstraction, relate them to each other, and combine them in beneficial ways.
            This applies across many disciplines, but the biggest dividing line that needs to be bridged is that between deductive and inductive approaches---or between explicit modelling of concepts and learning from data, and the trade-off between correctness and precision.  In the existing curricula (e.g., \cite{Kumar+24})
            abstraction occurs primarily in the form of abstract data types, abstraction in functional programming, and the difference between software architecture and design, but does not appear to be discussed, for example, in the scope of AI and data science. We propose to enhance existing curricula 
              with a cross-cutting perspective on abstraction.
      \item[How do we teach?]
            Teaching `abstraction' is difficult and effective methods to do so have been studied for some time (for an overview, see \cite{Mirolo+21}), not least under the---sometimes considered controversial---heading of ``computational thinking''.
            As we have seen, we will need to teach more: not just what abstraction is, but how to engineer abstractions useful in different disciplines and how to combine different approaches to abstraction engineering usefully.
            This will require further pedagogy and didactic research.
    \end{description}


  \subsection{Artifacts}
  
    There are several concrete products we need to deliver to advance Abstraction Engineering:
    \begin{description}
    
      \item[Abstraction Engineering Body of Knowledge:] 
            We call on the community to join us in building the AEBoK---\emph {Abstraction Engineering Body of Knowledge}. 
            AEBoK can start from cataloguing relevant existing techniques from the various disciplines that inspired the Abstraction Engineering field (e.g., model-driven engineering, modeling \& simulation, formal methods, AI, etc.).
            Beyond those, it should be an openly accessible document capturing the community-agreed terminology and core results in Abstraction Engineering.
            
      \item[Process Models and Tool Support:]
            As the community makes\\ progress with addressing the RQs and challenges, it will be important to develop process models and tool chains (including well-defined APIs) to facilitate the rigorous development, use, assurance of AE-based systems.
      
      \item[Training materials and tools:]
            In addition to research, tooling, validation studies, it is important to develop and disseminate educational materials that provide practitioners and students with theoretical foundations and hands-on experience with developing, analyzing, and maintaining AE artifacts with real-world and industrial-strength applications.
            These training materials and tools will underpin the education roadmap outlined in \cref{s:roadmap:education}.
    \end{description}      
  
  \subsection{Ways of working}
  
    Abstraction Engineering is inherently inter-disciplinary and depends on moving outside of historically established silos and research communities.
    This suggests certain ways of working that will be essential for successfully establishing Abstraction Engineering as a new principled lens on software and systems engineering:
    
    \begin{description}
      \item[Multidisciplinary Collaborations:]
            The majority of Abstraction Engineering projects should involve multidisciplinary expertise, including domain experts (e.g., legal, environmental, healthcare, energy, and etc.), safety engineers, AI/ML, data scientists, control theory, simulation, HCI, modeling, formal methods, cognitive psychology, and more.
            This will ensure the best possible transfer of knowledge across disciplines, allowing us to not only solve problems in specific silos but learn and generalise from different approaches taken in separate disciplines.
      \item[Partnerships with Industrial Collaborators:]
            This paper has\\ posed the challenges and RQs in terms of real-world systems that are becoming ubiquitous, with a general trend towards increasing automation.
            As such, any Abstraction Engineering research should be formally grounded, while being guided by industrial-strength challenge problems and real world operating contexts.
            Ideally, again, where problems from multiple domains can be brought together, it will become possible to understand how Abstraction Engineering approaches translate across disciplines.
    \end{description}
\section{Conclusions}
\label {sec:conclusion}

  In this paper, we discussed the challenges of systematic engineering of software systems in times when the boundary between the systems and the real world becomes increasingly blurred, and when the functional and non-functional requirements of such systems are increasingly difficult to strictly specify.
  More specifically, we explored challenges related to the fast-growing temporal adaptability in the development and evolution of all types of modern software systems.

  A common element in all adaptation strategies is the use of different types of abstractions.
  Abstractions are key to understanding, assuring and maintaining software systems.
  Mastering the engineering (i.e., construction and manipulation) of such abstractions is therefore essential to successfully creating future software systems.
  We called this process \emph {Abstraction Engineering}.
  In the spirit of Naur's statement that ``programming is theory building''\,\cite{Naur1986}, \emph {software engineering is Abstraction Engineering}.
  
  On the one hand, temporal adaptability pressures engineering methods into more flexible abstractions and into abstractions obtained and manipulated by machines.
  It is no longer possible to adapt abstractions solely through human-driven processes.
  On the other hand, our field cannot afford to give up understanding of the systems we engineer.
  Giving in to black-box end-to-end-learning designs~\cite{DBLP:journals/corr/BojarskiTDFFGJM16,pmlr-v70-silver17a}, where the interfaces between components do not carry explanatory semantics, is not the future for software engineering.
  Abstraction Engineering will be crucial to decompose, explain and control the design of future systems.
  Given the revolutionary role that software is playing for the modern day, compounded by the challenges posed by uncertainty and complexity, Abstraction Engineering will necessarily play a significant role in the revolution of software engineering to support the development and maintenance of modern software-based systems.

  As a first step on the road to such a renewed emphasis on abstractions and their engineering, we have highlighted the use of abstraction in several application domains, and extracted 13 fundamental research challenges from these examples. 
  On this foundation, we are proposing a roadmap for future research and education in the field of abstraction engineering.
  This roadmap will require efforts across communities and disciplines, overcoming the strict disciplinary silos we often encounter.

\begin{acks}
  Zschaler's work has been partly supported by the \grantsponsor{EPSRC}{UK Engineering and Physical Sciences Research Council (EPSRC)}{https://www.ukri.org/councils/epsrc/} under grant number \grantnum{EPSRC}{EP/T030747/1}.
\end{acks}

\bibliographystyle {plain}
\bibliography {biblio}

\begin{thebibliography}{10}

\bibitem{albarghouthi12}
Aws Albarghouthi, Arie Gurfinkel, and Marsha Chechik.
\newblock {From Under-Approximations to Over-Approximations and Back}.
\newblock In {\em Proc. of TACAS'2012}, pages 157--172, 2012.

\bibitem{alvi.ea:2023:deep}
Maira Alvi, Damien Batstone, Christian~Kazadi Mbamba, Philip Keymer, Tim
  French, Andrew Ward, Jason Dwyer, and Rachel Cardell-Oliver.
\newblock Deep learning in wastewater treatment: {A} critical review.
\newblock {\em Water Research}, 245:120518, 2023.

\bibitem{10.1145/1882362.1882367}
Luciano Baresi and Carlo Ghezzi.
\newblock The disappearing boundary between development-time and run-time.
\newblock In {\em Proceedings of the FSE/SDP Workshop on Future of Software
  Engineering Research}, FoSER '10, page 17–22, New York, NY, USA, 2010.
  Association for Computing Machinery.

\bibitem{viper}
Osbert Bastani, Yewen Pu, and Armando Solar{-}Lezama.
\newblock Verifiable reinforcement learning via policy extraction.
\newblock In Samy Bengio, Hanna~M. Wallach, Hugo Larochelle, Kristen Grauman,
  Nicol{\`{o}} Cesa{-}Bianchi, and Roman Garnett, editors, {\em Advances in
  Neural Information Processing Systems 31: Annual Conference on Neural
  Information Processing Systems 2018, NeurIPS 2018, December 3-8, 2018,
  Montr{\'{e}}al, Canada}, pages 2499--2509, 2018.

\bibitem{belfadil.ea:2024:leveraging}
Anas Belfadil, David Modesto, Jordi Meseguer, Bernat Joseph-Duran, David
  Saporta, and Jose Antonio~Martin Hernandez.
\newblock Leveraging deep reinforcement learning for water distribution systems
  with large action spaces and uncertainties: {DRL-EPANET} for pressure
  control.
\newblock {\em Journal of Water Resources Planning and Management},
  150(2):04023076, 2024.

\bibitem{Bencomo2013}
N.~Bencomo, A.~Bennaceur, P.~Grace, G.~Blair, and V.~Issarny.
\newblock The role of models@run.time in supporting on-the-fly
  interoperability.
\newblock {\em Computing}, 95(3):167–190, mar 2013.

\bibitem{BencomoGS19}
Nelly Bencomo, Sebastian G{\"{o}}tz, and Hui Song.
\newblock Models@run.time: a guided tour of the state of the art and research
  challenges.
\newblock {\em Softw. Syst. Model.}, 18(5):3049--3082, 2019.

\bibitem{reqatruntime}
Nelly Bencomo, Jon Whittle, Pete Sawyer, Anthony Finkelstein, and Emmanuel
  Letier.
\newblock Requirements reflection: requirements as runtime entities.
\newblock In {\em Proceedings of the 32nd ACM/IEEE International Conference on
  Software Engineering - Volume 2}, ICSE '10, page 199–202, New York, NY,
  USA, 2010. Association for Computing Machinery.

\bibitem{bennaceur19}
Amel Bennaceur, Andrea Zisman, Ciaran McCormick, Danny Barthaud, and Bashar
  Nuseibeh.
\newblock {Won't take no for an answer: resource-driven requirements
  adaptation}.
\newblock In {\em Proc. of SEAMS@ICSE'19}, pages 77--88, 2019.

\bibitem{DBLP:journals/corr/BojarskiTDFFGJM16}
Mariusz Bojarski, Davide~Del Testa, Daniel Dworakowski, Bernhard Firner, Beat
  Flepp, Prasoon Goyal, Lawrence~D. Jackel, Mathew Monfort, Urs Muller, Jiakai
  Zhang, Xin Zhang, Jake Zhao, and Karol Zieba.
\newblock End to end learning for self-driving cars.
\newblock {\em CoRR}, abs/1604.07316, 2016.

\bibitem{Brambilla+17}
Marco Brambilla, Jordi Cabot, Manuel Wimmer, and Luciano Baresi.
\newblock {\em Model-Driven Software Engineering in Practice}.
\newblock Morgan \& Claypool, second edition edition, 2017.

\bibitem{butting2020compositional}
Arvid Butting, Jerome Pfeiffer, Bernhard Rumpe, and Andreas Wortmann.
\newblock A compositional framework for systematic modeling language reuse.
\newblock In {\em Proceedings of the 23rd ACM/IEEE International Conference on
  Model Driven Engineering Languages and Systems}, pages 35--46, 2020.

\bibitem{calinescu17}
Radu Calinescu, Simos Gerasimou, Ibrahim Habli, M.~Iftikhar, Tim Kelly, and
  Danny Weyns.
\newblock Engineering trustworthy self-adaptive software with dynamic assurance
  cases.
\newblock {\em IEEE Transactions on Software Engineering}, PP, 03 2017.

\bibitem{camara2023assessment}
Javier C{\'a}mara, Javier Troya, Lola Burgue{\~n}o, and Antonio Vallecillo.
\newblock On the assessment of generative ai in modeling tasks: an experience
  report with {ChatGPT} and {UML}.
\newblock {\em Software and Systems Modeling}, pages 1--13, 2023.

\bibitem{DBLP:journals/sosym/CamaraTVBCCGS22}
Javier C{\'{a}}mara, Javier Troya, Antonio Vallecillo, Nelly Bencomo, Radu
  Calinescu, Betty H.~C. Cheng, David Garlan, and Bradley~R. Schmerl.
\newblock The uncertainty interaction problem in self-adaptive systems.
\newblock {\em Softw. Syst. Model.}, 21(4):1277--1294, 2022.

\bibitem{AC-ROS-MODELS2020}
Betty H.~C. Cheng, Robert~J. Clark, Jonathon~E. Fleck, Michael~A. Langford, and
  Philip~K. McKinley.
\newblock {AC-ROS: Assurance Case Driven Adaptation for the Robot Operating
  System}.
\newblock In {\em {Proc. 23rd Int. Conf. on Model Driven Engineering Languages
  and Systems (MODELS 2020)}}, pages 102--113. ACM, 2020.

\bibitem{Clarke+00}
Edmund Clarke, Orna Grumberg, Somesh Jha, Yuan Lu, and Helmut Veith.
\newblock Counterexample-guided abstraction refinement.
\newblock In E.~Allen Emerson and Aravinda~Prasad Sistla, editors, {\em Proc.
  12th Int'l Conf. Computer Aided Verification (CAV'00)}, pages 154--169.
  Springer Berlin Heidelberg, 2000.

\bibitem{mdebook}
Benoit Combemale, Robert France, Jean-Marc Jézéquel, Bernhard Rumpe, Jim
  Steel, and Didier Vojtisek.
\newblock {\em Engineering Modeling Languages: Turning Domain Knowledge into
  Tools}.
\newblock Taylor \& Francis, 2016.

\bibitem{Combemale+21DS}
Benoit Combemale, Jorg Kienzle, Gunter Mussbacher, Hyacinth Ali, Daniel Amyot,
  Mojtaba Bagherzadeh, Edouard Batot, Nelly Bencomo, Benjamin Benni,
  Jean-Michel Bruel, Jordi Cabot, Betty~H.C. Cheng, Philippe Collet, Gregor
  Engels, Robert Heinrich, Jean-Marc Jezequel, Anne Koziolek, Sebastien Mosser,
  Ralf Reussner, Houari Sahraoui, Rijul Saini, June Sallou, Serge Stinckwich,
  Eugene Syriani, and Manuel Wimmer.
\newblock A hitchhiker's guide to model-driven engineering for data-centric
  systems.
\newblock {\em {IEEE} Software}, 38(4):71--84, 2021.

\bibitem{CousotCousot77}
Patrick Cousot and Radhia Cousot.
\newblock Abstract interpretation: a unified lattice model for static analysis
  of programs by construction or approximation of fixpoints.
\newblock In {\em Proceedings of the 4th ACM SIGACT-SIGPLAN Symposium on
  Principles of Programming Languages}, POPL '77, page 238–252, New York, NY,
  USA, 1977. Association for Computing Machinery.

\bibitem{Dalibor+22}
Manuela Dalibor, Nico Jansen, Bernhard Rumpe, David Schmalzing, Louis
  Wachtmeister, Manuel Wimmer, and Andreas Wortmann.
\newblock A cross-domain systematic mapping study on software engineering for
  digital twins.
\newblock {\em Journal of Systems and Software}, 2022.

\bibitem{dippolito08}
N.~D'Ippolito, D.~Fischbein, M.~Chechik, and S.~Uchitel.
\newblock {MTSA: The Modal Transition System Analyser}.
\newblock In {\em Proc. of 23rd IEEE/ACM International Conference on Automated
  Software Engineering}, pages 475--476, 2008.

\bibitem{DTConcepts}
Romina Eramo, Francis Bordeleau, Benoit Combemale, Mark van~den Brand, Manuel
  Wimmer, and Andreas Wortmann.
\newblock Conceptualizing digital twins.
\newblock {\em IEEE Software}, 39(2):39--46, 2022.

\bibitem{10.1145/2427048.2427055}
Sebastian Erdweg, Paolo~G. Giarrusso, and Tillmann Rendel.
\newblock Language composition untangled.
\newblock In {\em Proceedings of the Twelfth Workshop on Language Descriptions,
  Tools, and Applications}, LDTA '12, New York, NY, USA, 2012. Association for
  Computing Machinery.

\bibitem{fill2023conceptual}
Hans-Georg Fill, Peter Fettke, and Julius K{\"o}pke.
\newblock Conceptual modeling and large language models: impressions from first
  experiments with {ChatGPT}.
\newblock {\em Enterprise Modelling and Information Systems Architectures
  (EMISAJ)}, 18:1--15, 2023.

\bibitem{AdaptiveTesting-SEAMS2014}
Erik~M. Fredericks, Byron DeVries, and Betty H.~C. Cheng.
\newblock Towards run-time adaptation of test cases for self-adaptive systems
  in the face of uncertainty.
\newblock In Gregor Engels and Nelly Bencomo, editors, {\em 9th International
  Symposium on Software Engineering for Adaptive and Self-Managing Systems,
  {SEAMS} 2014, Proceedings, Hyderabad, India, June 2-3, 2014}, pages 17--26.
  {ACM}, 2014.

\bibitem{fujiyoshi.ea:2019:deep}
Hironobu Fujiyoshi, Tsubasa Hirakawa, and Takayoshi Yamashita.
\newblock Deep learning-based image recognition for autonomous driving.
\newblock {\em IATSS Research}, 43(4):244--252, 2019.

\bibitem{gerhold2018model}
Marcus Gerhold and Mari{\"e}lle Stoelinga.
\newblock Model-based testing of probabilistic systems.
\newblock {\em Formal aspects of computing}, 30:77--106, 2018.

\bibitem{gray.rumpe:2022}
J.~Gray and B.~Rumpe.
\newblock Modeling of, for, and with digital twins.
\newblock {\em Software and Systems Modeling}, 21:1685--1686, 2022.

\bibitem{Guizzardi05}
Giancarlo Guizzardi.
\newblock {\em Ontological foundations for structural conceptual models}.
\newblock {PhD} thesis, University of Twente, 2005.

\bibitem{ParnasBook2001}
Daniel~M. Hoffman and David~M. Weiss, editors.
\newblock {\em Software fundamentals: collected papers by David L. Parnas}.
\newblock Addison-Wesley Longman Publishing Co., Inc., USA, 2001.

\bibitem{hu22}
Boyue~Caroline Hu, Lina Marsso, Krzysztof Czarnecki, Rick Salay, Huakun Shen,
  and Marsha Chechik.
\newblock {If a Human Can See It, So Should Your System: Reliability
  Requirements for Machine Vision Components}.
\newblock In {\em Proceedings of the 44th International Conference on Software
  Engineering (ICSE'2022), Pittsburgh, USA}. {ACM}, 2022.

\bibitem{hu23}
Boyue~Caroline Hu, Lina Marsso, Nikita Dvornik, Huakun Shen, and Marsha
  Chechik.
\newblock {DecompoVision: Reliability Analysis of Machine Vision Components
  through Decomposition and Reuse}.
\newblock In {\em Proceedings of the 31st {ACM} Joint European Software
  Engineering Conference and Symposium on the Foundations of Software
  Engineering, {ESEC/FSE} 2023}, pages 541--552. {ACM}, 2023.

\bibitem{ISO21448:2022}
{ISO 21448:2022, Road vehicles -- Safety of the intended functionality}.
\newblock International Standard, 6 2022.
\newblock International Standard published [60.60], Technical Committee: ISO/TC
  22/SC 32.

\bibitem{ivanov.ea:2020}
Sergey Ivanov, Ksenia Nikolskaya, Gleb Radchenko, Leonid Sokolinsky, and
  Mikhail Zymbler.
\newblock Digital twin of city: Concept overview.
\newblock In {\em 2020 Global Smart Industry Conference (GloSIC)}, pages
  178--186, 2020.

\bibitem{Josang01}
Audun J{\o}sang.
\newblock A logic for uncertain probabilities.
\newblock {\em International Journal of Uncertainty, Fuzziness and
  Knowledge-Based Systems}, 9(3):279--212, 2001.

\bibitem{Kang+90}
Kyo Kang, Sholom Cohen, James Hess, William Novak, and Spencer Peterson.
\newblock Feature-oriented domain analysis {(FODA)} feasibility study.
\newblock Technical Report CMU/SEI-90-TR-0211990, Software Engineering
  Institute, 1990.

\bibitem{Kiczales1991}
G.~Kiczales.
\newblock Towards a new model of abstraction in software engineering.
\newblock In {\em Proceedings 1991 International Workshop on Object Orientation
  in Operating Systems}, pages 127--128, 1991.

\bibitem{10.5555/1496375}
Anneke Kleppe.
\newblock {\em Software Language Engineering: Creating Domain-Specific
  Languages Using Metamodels}.
\newblock Addison-Wesley Professional, 1 edition, 2008.

\bibitem{Kramer07}
Jeff Kramer.
\newblock Is abstraction the key to computing?
\newblock {\em Commun. ACM}, 50(4):36--42, April 2007.

\bibitem{kramer2006role}
Jeff Kramer and Orit Hazzan.
\newblock Summary of an {ICSE} 2006 workshop: The role of abstraction in
  software engineering.
\newblock {\em ACM SIGSOFT Software Engineering Notes}, 31(6):38--39, 2006.

\bibitem{Kumar+24}
Amruth~N. Kumar, Rajendra~K. Raj, Sherif~G. Aly, Monica~D. Anderson, Brett~A.
  Becker, Richard~L. Blumenthal, Eric Eaton, Susan~L. Epstein, Michael
  Goldweber, Pankaj Jalote, Douglas Lea, Michael Oudshoorn, Marcelo Pias, Susan
  Reiser, Christian Servin, Rahul Simha, Titus Winters, and Qiao Xiang.
\newblock {\em Computer Science Curricula 2023}.
\newblock Association for Computing Machinery, New York, NY, USA, 2024.

\bibitem{Enlil-SEAMS2021}
Michael~A. Langford and Betty H.~C. Cheng.
\newblock {``Know What You Know'': Predicting Behavior for Learning-Enabled
  Systems When Facing Uncertainty}.
\newblock In {\em {Proc. 16th Int. Symposium on Software Engineering for
  Adaptive and Self-Managing Systems (SEAMS 2021)}}, pages 78--89. ACM, 2021.

\bibitem{MODELS2022-MODALAS-MDE-Assurance-LearningBasedSystems}
Michael~Austin Langford, Kenneth~H. Chan, Jonathon~Emil Fleck, Philip~K.
  McKinley, and Betty~H.C. Cheng.
\newblock Modalas: Model-driven assurance for learning-enabled autonomous
  systems.
\newblock In {\em 2021 ACM/IEEE 24th International Conference on Model Driven
  Engineering Languages and Systems (MODELS)}, pages 182--193, 2021.

\bibitem{Enki-TAAS2021}
Michael~Austin Langford and Betty H.~C. Cheng.
\newblock {Enki: A Diversity-Driven Approach to Test and Train Robust
  Learning-Enabled Systems}.
\newblock {\em ACM Trans. Auton. Adapt. Syst.}, 15(2), May 2021.

\bibitem{Annunaki-TAAS2024}
Michael~Austin Langford, Sol Zilberman, and Betty~H.C. Cheng.
\newblock Anunnaki: A modular framework for developing trusted artificial
  intelligence.
\newblock {\em ACM Trans. Auton. Adapt. Syst.}, March 2024.
\newblock (in press).

\bibitem{lefevre.ea:2016:tase}
Stéphanie Lefevre, Ashwin Carvalho, and Francesco Borrelli.
\newblock A learning-based framework for velocity control in autonomous
  driving.
\newblock {\em IEEE Trans Autom. Sci. Eng.}, 13(1):32--42, 2016.

\bibitem{prism}
D.~Parker M.~Kwiatkowska, G.~Norman.
\newblock {PRISM 4.0: Verification of Probabilistic Real-time Systems}.
\newblock In {\em Proc. 23rd International Conference on Computer Aided
  Verification (CAV’11)}, volume 6806 of {\em Lecture Notes in Computer
  Science}, pages 585--591. Springer, 2011.

\bibitem{Mirolo+21}
Claudio Mirolo, Cruz Izu, Violetta Lonati, and Emanuele Scapin.
\newblock Abstraction in computer science education: An overview.
\newblock {\em Informatics in Education}, 20(4):615--639, December 2021.

\bibitem{DBLP:conf/models/MussbacherABBCCCFHHKSSSW14}
Gunter Mussbacher, Daniel Amyot, Ruth Breu, Jean{-}Michel Bruel, Betty H.~C.
  Cheng, Philippe Collet, Beno{\^{\i}}t Combemale, Robert~B. France, Rogardt
  Heldal, James~H. Hill, J{\"{o}}rg Kienzle, Matthias Sch{\"{o}}ttle, Friedrich
  Steimann, Dave~R. Stikkolorum, and Jon Whittle.
\newblock The relevance of model-driven engineering thirty years from now.
\newblock In {\em Model-Driven Engineering Languages and Systems - 17th
  International Conference, {MODELS} 2014, Valencia, Spain, September 28 -
  October 3, 2014. Proceedings}, volume 8767 of {\em Lecture Notes in Computer
  Science}, pages 183--200. Springer, 2014.

\bibitem{Naur1986}
Peter Naur.
\newblock Programming as theory building.
\newblock {\em Microprocessing and Microprogramming}, 15:253--261, 1986.

\bibitem{Niederer2021}
Steven~A Niederer, Michael~S Sacks, Mark Girolami, and Karen Willcox.
\newblock Scaling digital twins from the artisanal to the industrial.
\newblock {\em Nature Computational Science}, 1(5):313--320, 2021.

\bibitem{DBLP:journals/re/SalayCHS13}
Rick Salay, Marsha Chechik, Jennifer Horkoff, and Alessio~Di Sandro.
\newblock Managing requirements uncertainty with partial models.
\newblock {\em Requir. Eng.}, 18(2):107--128, 2013.

\bibitem{schrotter.ea:2020:digital}
Gerhard Schrotter and Christian H{\"u}rzeler.
\newblock The digital twin of the city of {Zurich} for urban planning.
\newblock {\em PFG--Journal of Photogrammetry, Remote Sensing and
  Geoinformation Science}, 88(1):99--112, 2020.

\bibitem{schwarting.ea:2018:planning}
Wilko Schwarting, Javier Alonso-Mora, and Daniela Rus.
\newblock Planning and decision-making for autonomous vehicles.
\newblock {\em Annual Review of Control, Robotics, and Autonomous Systems},
  1:187--210, 2018.

\bibitem{sepulveda.ea:2018}
Gabriel Sepulveda, Juan~Carlos Niebles, and Alvaro Soto.
\newblock A deep learning based behavioral approach to indoor autonomous
  navigation.
\newblock {\em CoRR}, abs/1803.04119, 2018.

\bibitem{shalevshwartz.ea:2018:formal}
Shai Shalev-Shwartz, Shaked Shammah, and Amnon Shashua.
\newblock On a formal model of safe and scalable self-driving cars, 2018.

\bibitem{shein:2020:zdnet}
Esther Shein.
\newblock How {AI} and {ML} are helping first responders. artificial
  intelligence and machine learning are helping firefighters, police officers,
  and {EMS} medics respond to emergencies and save lives.
\newblock {\em ZDNet}, 2020.
\newblock Accessed: 2024-03-25.

\bibitem{pmlr-v70-silver17a}
David Silver, Hado van Hasselt, Matteo Hessel, Tom Schaul, Arthur Guez, Tim
  Harley, Gabriel Dulac-Arnold, David Reichert, Neil Rabinowitz, Andre Barreto,
  and Thomas Degris.
\newblock The predictron: End-to-end learning and planning.
\newblock In Doina Precup and Yee~Whye Teh, editors, {\em Proceedings of the
  34th International Conference on Machine Learning}, volume~70 of {\em
  Proceedings of Machine Learning Research}, pages 3191--3199. PMLR, 06--11 Aug
  2017.

\bibitem{tampuu.ea:2022}
Ardi Tampuu, Tambet Matiisen, Maksym Semikin, Dmytro Fishman, and Naveed
  Muhammad.
\newblock A survey of end-to-end driving: Architectures and training methods.
\newblock {\em IEEE Transactions on Neural Networks and Learning Systems},
  33(4):1364--1384, 2022.

\bibitem{tenorth.beetz:2009:knowrob}
Moritz Tenorth and Michael Beetz.
\newblock {KNOWROB}: knowledge processing for autonomous personal robots.
\newblock In {\em IROS}, pages 4261--4266. IEEE, 2009.

\bibitem{thrun:2005:probabilistic}
Sebastian Thrun, Wolfram Burgard, and Dieter Fox.
\newblock {\em Probabilistic robotics}.
\newblock MIT Press, Cambridge, Mass., 2005.

\bibitem{DBLP:journals/pacmpl/VarshosazGJW23}
Mahsa Varshosaz, Mohsen Ghaffari, Einar~Broch Johnsen, and Andrzej Wąsowski.
\newblock Formal specification and testing for reinforcement learning.
\newblock {\em Proc. {ACM} Program. Lang.}, 7({ICFP}):125--158, 2023.

\bibitem{Visser15}
Eelco Visser.
\newblock Understanding software through linguistic abstraction.
\newblock {\em Science of Computer Programming}, 97:11--16, 2015.
\newblock Special Issue on New Ideas and Emerging Results in Understanding
  Software.

\bibitem{wang.ea:2023:fse}
Longtian Wang, Xiaofei Xie, Xiaoning Du, Meng Tian, Qing Guo, Zheng Yang, and
  Chao Shen.
\newblock Distxplore: Distribution-guided testing for evaluating and enhancing
  deep learning systems.
\newblock In {\em Proceedings of the 31st ACM Joint European Software
  Engineering Conference and Symposium on the Foundations of Software
  Engineering}, ESEC/FSE 2023, page 68–80, New York, NY, USA, 2023.
  Association for Computing Machinery.

\bibitem{Ward1994}
M.~P. Ward.
\newblock Language-oriented programming.
\newblock {\em Software-Concepts and Tools}, 15(4):147--161, 1994.

\bibitem{10.1145/3485275}
Cody Watson, Nathan Cooper, David~Nader Palacio, Kevin Moran, and Denys
  Poshyvanyk.
\newblock A systematic literature review on the use of deep learning in
  software engineering research.
\newblock {\em ACM Trans. Softw. Eng. Methodol.}, 31(2), mar 2022.

\bibitem{Wei+19}
Ran Wei, Tim~P. Kelly, Xiaotian Dai, Shuai Zhao, and Richard Hawkins.
\newblock Model based system assurance using the structured assurance case
  metamodel.
\newblock {\em Journal of Systems and Software}, 154:211--233, 2019.

\bibitem{weyns2021towards}
Danny Weyns, Bradley Schmerl, Masako Kishida, Alberto Leva, Marin Litoiu,
  Necmiye Ozay, Colin Paterson, and Kenji Tei.
\newblock Towards better adaptive systems by combining mape, control theory,
  and machine learning.
\newblock In {\em 2021 International Symposium on Software Engineering for
  Adaptive and Self-Managing Systems (SEAMS)}, pages 217--223. IEEE, 2021.

\bibitem{dsldesign}
Andrzej Wąsowski and Thorsten Berger.
\newblock {\em Domain-Specific Languages. Effective Modeling, Automation, and
  Reuse}.
\newblock Springer, 2023.

\bibitem{10.1145/3505243}
Yanming Yang, Xin Xia, David Lo, and John Grundy.
\newblock A survey on deep learning for software engineering.
\newblock {\em ACM Comput. Surv.}, 54(10s), sep 2022.

\end{thebibliography}

\end {document}